# Velocity tuning of friction with two trapped atoms

Dorian Gangloff[1,†], Alexei Bylinskii[1,†], Ian Counts[1], Wonho Jhe[2], and Vladan Vuletić[1,*]

[1]Department of Physics, and Research Laboratory of Electronics, Massachusetts Institute of Technology, Cambridge, MA, USA
[2]Department of Physics and Astronomy, Seoul National University, Seoul, Korea
[†]these authors contributed equally
*vuletic@mit.edu

**Friction is the basic, ubiquitous mechanical interaction between two surfaces that results in resistance to motion and energy dissipation. In spite of its technological and economic significance, our ability to control friction remains modest, and our understanding of the microscopic processes incomplete[1–3]. At the atomic scale, mismatch between the two contacting crystal lattices can lead to a reduction of stick-slip friction (structural lubricity)[4–7], while thermally activated atomic motion can give rise to a complex velocity dependence[8–14], and nearly vanishing friction at sufficiently low velocities (thermal lubricity)[10,13]. Atomic force microscopy has provided a wealth of experimental results[6–9,14–19], but limitations in the dynamic range, time resolution, and control at the single-atom level have hampered a full quantitative description from first principles[3]. Here, using an ion-crystal friction emulator with single-atom, single substrate-site spatial resolution and single-slip temporal resolution[5,20], we measure the friction force over nearly five orders of magnitude in velocity, and contiguously observe four distinct regimes, while controlling temperature and dissipation. We elucidate the interplay between thermal and structural lubricity in a system of two coupled atoms, and provide a simple explanation in terms of the Peierls-Nabarro potential[21]. This extensive control at the atomic scale paves the way for fundamental studies of the interaction of many-atom surfaces, as for example in the Frenkel-Kontorova model[21], and possibly into the quantum regime[22].**

In the simplest scenario for stick-slip friction, a single atom at an object-substrate interface experiences a force resisting its motion due to a periodic potential created by the substrate[2,3,13,23] (Fig. 1a). A finite external force is then required to pull the atom from one potential well and cause it to slip to the next well across an energy barrier $U_B$. Interestingly, in the case of more than one atom forming the contact interface, friction can be dramatically reduced by structural mismatch of the object and substrate, an effect coined superlubricity[4–7], and observed in friction force miscroscopy[7,17], colloidal monolayers[24], and in our friction emulator recently[5]. Thermally activated transitions between neighboring potential wells at temperatures $T \sim U_B$ can also reduce the friction force significantly, making it velocity-dependent[8–14]. Observations in different systems have spanned from the high-temperature regime of thermolubricity[10] to the low-temperature regime of strong stick-slip[9]. In the present work, as a function of velocity, we observe the continuous transition between the regimes of thermal drift, where friction is small and (nearly) velocity independent, thermal activation, where friction increases logarithmically with velocity, stick-slip, where friction is large and nearly velocity independent, and underdamping, where friction decreases with velocity because the damping is not fast enough to remove the energy released in a slip. For a two-atom contact, we observe that the measured friction force is substantially reduced by the interaction between the atoms when they are arranged so as to cancel the forces from the substrate[5]. We link this structural lubricity to a reduced barrier $\widetilde{U}_B < U_B$ in the Peierls-Nabarro potential[21,25,26], and distinguish structurally-induced thermolubricity ($T \sim \widetilde{U}_B$) from structural lubricity ($T \ll \widetilde{U}_B$) by observing the velocity-dependence.

Single-asperity friction experiments have been performed with an atomically sharp tip that is translated across an atomically smooth surface[16]. Our emulator[5,20,27] of nanofriction[26,28] consists of one or two electrically-trapped atomic ions[29] pulled against the sinusoidal potential (Fig. 1a,b) of a standing wave of light (optical lattice)[20,30,31]. We observe each ion's trajectory with resolution below the period of the optical lattice via the ion's position-dependent fluorescence[5]. Each time the ion slips into the next lattice well, its fluorescence reaches a maximum and decreases as the ion is laser-cooled into the new potential minimum (Fig. 1c,d). Hysteresis in the timing of the slip as the electrostatic parabolic trapping potential is pulled back and forth reveals the maximum static friction force exerted by the lattice on the ion (Fig. 1d). At finite ion temperature $T$, the observed hysteresis and corresponding friction force are reduced (Fig. 1c,d).

At zero temperature[2,11,23], the dynamical behavior is completely determined by the ratio of the lattice confinement frequency $\omega_l \propto \sqrt{U_l}$ to the electrostatic confinement frequency $\omega_0 \propto \sqrt{K}$ (Fig. 1a), where $U_l$ is the depth of the lattice potential, and $K$ is the spring constant of the electrostatic trap. The corresponding dimensionless corrugation parameter $\eta = \omega_l^2/\omega_0^2$ determines the number of minima in the overall potential energy landscape experienced by a single ion. For $\eta \leq 1$, there is only a single minimum that is translated with the applied force, and there is no stick-slip friction. In our regime of interest, $1 < \eta < 4.6$, the confinements exerted by the lattice and the electrostatic trap are comparable, and there are at most two local minima in the overall potential at any time, separated by a maximum energy barrier $U_B/U_l \simeq (\eta - 1)^2/\eta^2$. At finite temperature, the ion can also slip via thermal activation before the barrier height is reduced to zero by the applied force (Fig. 1c), leading to a reduced hysteresis, and friction (thermolubricity) that depends on the transport velocity $v$.

For the first time contiguously in a single experiment, we observe for a single ion four regimes of friction with distinct velocity dependences (Fig. 2a). These regimes can be organized by the hierarchy of three time scales, namely the thermal hopping time between lattice wells $\tau_{th}$, the transport time for the external trap to move by one lattice well $a/v$, and the ion recooling time $\tau_c$. When $\tau_{th} \ll a/v$, thermal hopping dominates, and the ion remains in thermal equilibrium, following the slowly moving ion trap – a regime called *thermal drift* where the friction force due to stick-slip (almost) vanishes[10]. In the thermal activation regime $\tau_{th} \sim a/v$, mechanical sticking to the barrier occurs often and contributes to an average friction force, which grows logarithmically with velocity[8,9,11,13]. For even larger velocities $\tau_{th} \gg a/v \gg \tau_c$, thermal hopping across lattice wells is negligible on the transport time scale $a/v$. This is the strong *stick-slip* regime where the friction force reaches its maximum value, present for a window of velocities where the atom reaches thermal equilibrium within a lattice well, but not between neighboring lattice wells. We also observe a fourth regime of friction, which we call *underdamping*, where the friction force decreases logarithmically with velocity, associated in solid-state systems with capillary condensation[18] and multi-asperity contacts[14]. In our system, this "velocity weakening" arises because the ion does not have sufficient time to recool after the slip for $a/v \lesssim \tau_c$. This effectively increases the ion's kinetic energy before the next slip event and reduces the friction force (Fig. 2a). Having direct access to all system parameters through independent microscopic measurements, we also show a full-dynamics simulation[11,13], without any free parameters, that closely follows our data over all four regimes of friction. Figure 2a furthermore shows that in the thermally activated and underdamped regimes, simple analytical models for the velocity-dependent friction developed previously[3,13] match our data quantitatively (Supplementary Information). The same good agreement between experimental data and theoretical models is attained when we change the corrugation depth $U_B$ or the temperature $T$ (Fig. 2b).

The friction force is expected to be particularly sensitive to temperature when $\tau_{th} \lesssim a/v$ due to exponential activation[10], and almost independent of it when $\tau_{th} \gg a/v$. In Fig. 3, we verify experimentally[15,19] that for low velocities ($\tau_{th} \lesssim a/v$) the friction force changes by an order of magnitude when we change the temperature by a factor of 6 (Supplementary Information), while for high velocities ($\tau_{th} \gg a/v$) the force varies by less than a factor of 2. This confirms that an effectively zero-temperature stick-slip regime[5] can be experimentally accessed at high transport velocity $v \gg a/\tau_{th}$.

In order to study the interplay between structural lubricity, arising from mismatch between the object and substrate corrugations, and thermolubricity, we place a second ion in the trap along the optical lattice direction (Fig. 1a,b). If the effective spring force arising from the Coulomb interaction between the ions were infinitely stiff, the friction force on the two-ion system could be made to vanish by placing the two ions at positions where they experience opposite lattice forces. It is the essence of structural lubricity that a substantial reduction in the friction force persists even for finite ion-ion interaction that is comparable to the substrate corrugation. When the ions experience opposite lattice forces, it is energetically favorable for them to pass the energy barrier between wells one at a time, as illustrated by the two-dimensional energy landscapes of Fig. 4d,e. This results in a reduced barrier depth $\widetilde{U}_B < U_B$ seen by each ion, and therefore a reduced friction force. Using the electrical trap, the spacing $d$ between the ions can be tuned to be an exact multiple of $a$ (i.e. $d \bmod a = 0$), or to be mismatched (i.e. $d \bmod a = a/2$). We have found in our previous work that mismatch causes a dramatic reduction of the observed friction force[5], as has been also observed for graphite flakes on a graphite substrate under certain orientations[7]. The friction reduction can be due to pure structural lubricity (stick-slip motion in a Peierls-Nabarro potential[21] with reduced energy barrier $\widetilde{U}_B \gg k_B T$) or to structurally-induced thermolubricity ($\widetilde{U}_B \sim k_B T$), here distinguished experimentally because only the latter is velocity-dependent.

In the matched case, the two-ion system is expected to behave as a rigid object akin to a single particle, because only the center-of-mass mode is affected by lattice forces. Figure 4a shows that the observed velocity-dependence of friction in the matched case indeed agrees with the one-ion case.

In the mismatched case, the lattice forces on the center-of-mass mode cancel out, and we observe that, for the same temperature, friction in the mismatched case is significantly reduced compared to the matched case (Fig. 4b), in good agreement with Langevin simulations. When comparing the friction in the mismatched case to the matched case, we find that there is no reduction in the thermal drift regime, and reduction by a factor of ~4.8 in the stick-slip regime (Fig. 4c). A zero-temperature evaluation of the two-ion energy landscape (Fig. 4d,e) shows that in the matched case, the barrier is identical to the one-ion case while in the mismatched case it is approximately four times lower ($\widetilde{U}_B/U_B \simeq 3.7$). The additional ~20% friction reduction observed compared to $\widetilde{U}_B/U_B$ can be explained by structurally-induced thermolubricity at fixed temperature due to the lower barrier depth $\widetilde{U}_B$ (Fig. 3). The high-velocity friction reduction plateau of Fig. 4c, where thermal hopping is negligible, then represents a direct observation of structurally-induced lubricity or "superlubricity"[4–7]. This interpretation is consistent with the observation that in this regime, the ions pass the barrier one at a time (Fig. 4e inset), reminiscent of a kink defect being transported across the two-atom chain[21]. Thus measuring the reduced friction force directly reveals the Peierls-Nabarro barrier[21] $\widetilde{U}_B$ for two atoms in a periodic potential.

The broad dynamic range of control and measurement demonstrated in this system enables the direct quantitative study of fundamental microscopic processes in stick-slip friction. In the future, the system could be used to study many-body phenomena arising from the strong particle interactions in the corrugated potential, such as the Aubry transition[2,21,26], while cooling to the vibrational ground state in a reduced potential may provide access to a regime of quantum friction dominated by quantum tunneling[22].

**Acknowledgements** We acknowledge support from the NSF-funded Center for Ultracold Atoms. D.G. and A.B. acknowledge funding from NSERC.

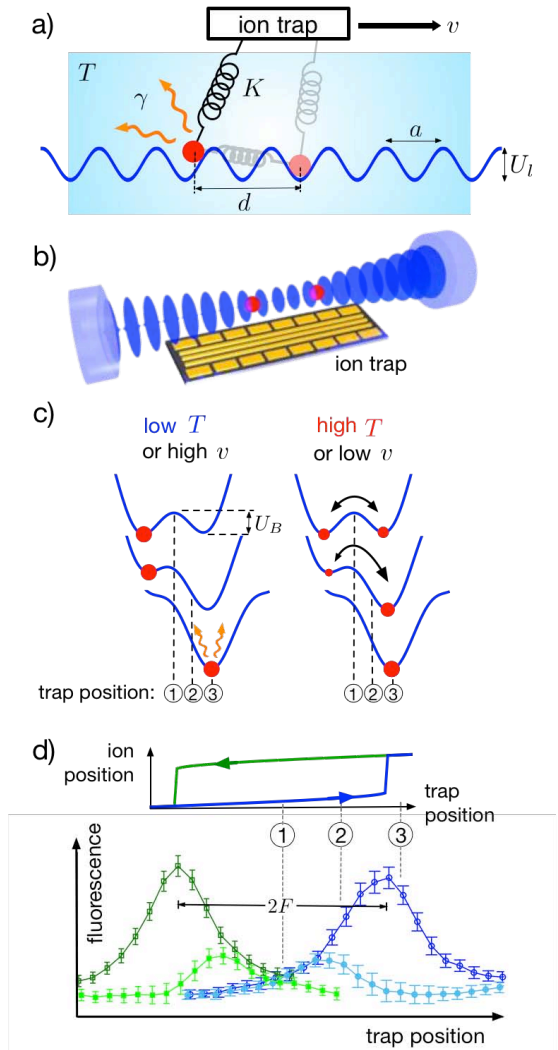

**Figure 1 | Friction emulator with trapped atomic ions in an optical lattice. a,b,** Model of the friction interface; one Yb$^+$ ion of mass $m = 2.9\times10^{-25}$ kg (or two coupled ions separated by a distance $d \approx 5$ μm) is confined in a Paul trap[27,29] with a spring constant $K = m\omega_0^2$ ($\omega_0/2\pi = 363$ kHz), whose equilibrium point is translated at a constant velocity $v$ by applying a time-varying electric field. An optical standing wave, detuned by ~ 12 GHz from the atomic $^2S_{1/2} \to\ ^2P_{1/2}$ transition, creates a sinusoidal potential of periodicity $a = 185$ nm and depth $U_l/h \approx 20$ MHz along the radiofrequency nodal line of the Paul trap[20]. The ion is kept at a temperature $T \approx 25$ μK via continuous laser cooling with a dissipation rate constant $\gamma = \tau_c^{-1} \approx 10^4\ s^{-1}$. **c, d,** Temperature dependence of stick-slip friction. At low temperature or high velocity, the ion sticks in its initial well, corresponding to a rise in its scattered fluorescence (1-2, blue open circles) until it slips to the next well, and the added energy is dissipated via laser cooling (3, blue open circles). The ion fluorescence is highest when the slip occurs. At high temperature or low velocity, the ion thermalizes over the energy barrier (1-2), and so smoothly transitions to the next well without frictional dissipation (3). (**d**) If the trap translation direction is reversed, maximum hysteresis is observed in the low-temperature or high-velocity regime (open symbols). A reduced hysteresis is present for an intermediate temperature or velocity regime (filled symbols). The friction force $F$ is measured via the separation $2F$ between the slips in the hysteresis loop[5]. In this and the following figures, error bars are statistical and represent one standard deviation.

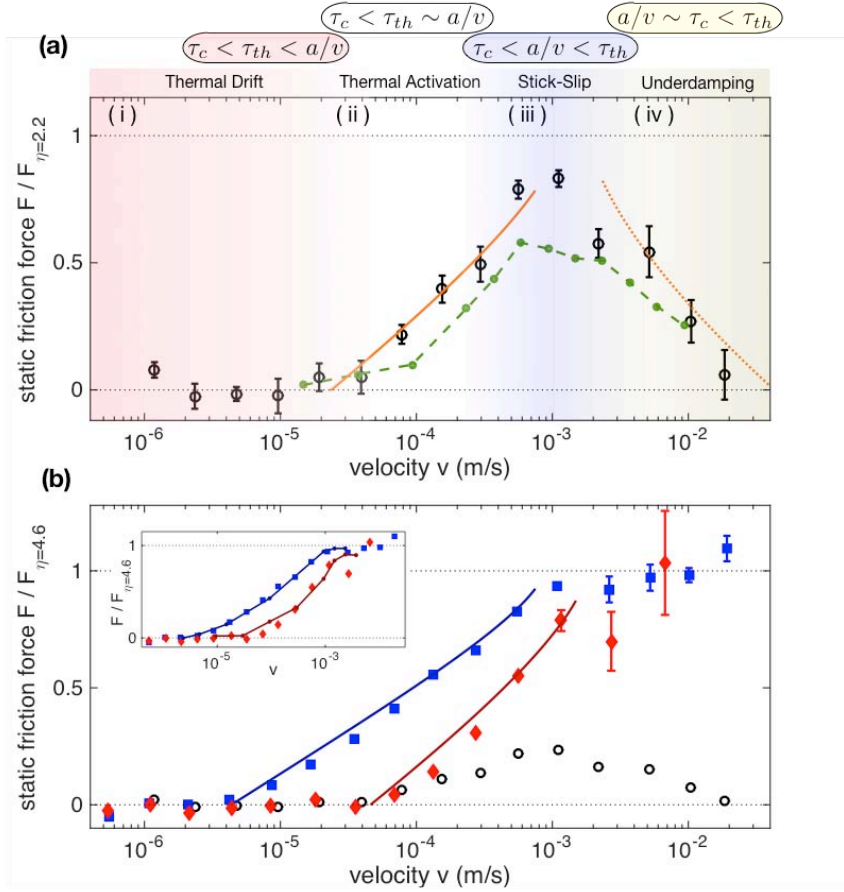

**Figure 2 | Velocity dependence of friction for one atom.** The transport time $a/v$ should be compared to two time scales: the thermal hopping time between two lattice wells, given by $\tau_{th} = \tau_0 \exp(U_B/k_B T)$ for a maximum barrier height $U_B$ (where $k_B$ is the Boltzmann constant and $\tau_0(\tau_c, \omega_0, \omega_l)$ is the hopping attempt time[12]); and the recooling time after a slip $\tau_c$. **a,** Here $\tau_{th} \approx 10\ ms$ and $\tau_c \approx 100\ \mu s$. Four regimes of friction are observed. $\eta = 2.2$ and the friction force is normalized by its zero-temperature maximum value $F_{\eta=2.2} \approx 0.17\ \pi U_l/a$. Here $U_l/h = 9.5$ MHz and $k_B T/U_l = 0.15(4)$. The solid orange line shows the expected result $F/F_{\eta=2.2} = 1 - \left(\frac{3}{2\sqrt{2}}\frac{k_B T}{U_l}\log\left(\frac{v_{th}}{v}\right)\right)^{2/3}$, where $v_{th} \sim 1$ mm/s, from an analytical model in the thermally activated regime[3,13]. Similarly in the underdamped regime, we model the friction as $F/F_{\eta=2.2} = 1 - \left(\frac{3}{2\sqrt{2}}\frac{k_B T}{U_l}\log\left(\frac{v}{v_c}\right)\right)^{2/3}$ (orange dotted line), where $k_B T/U_l = 0.3$ and $v_c = \frac{a}{\tau_c} \sim 2$ mm/s (Supplementary Information). The Langevin simulation (dashed green line) is in good agreement with the data over all four velocity regimes for parameters $\eta = 2.2$, $k_B T/U_l = 0.15$, $\tau_c = 100\ \mu s$. **b,** At a larger lattice depth $U_l/h = 20$ MHz where $\eta = 4.6$, increasing the temperature from $k_B T/U_l = 0.04(1)$ (blue squares) to $k_B T/U_l = 0.17(1)$ (red diamonds) reduces the friction in the thermally activated region $10^{-5}$ m/s $\lesssim v \lesssim 10^{-3}$ m/s while leaving stick-slip friction in the region $10^{-3}$ m/s $\lesssim v \lesssim 10^{-2}$ m/s almost unaffected. Here, $\tau_c \approx 50\ \mu s$. The friction force is normalized by its zero-temperature maximum value for $\eta = 4.6$, $F_{\eta=4.6} \approx 0.61\ \pi U_l/a$. Solid lines show the expected results from the analytical thermal activation model[3,13]. Data from (a), normalized to $F_{\eta=4.6}$, is shown as open black circles. Langevin simulations (inset, solid lines) are in good agreement with the data for parameters $\eta = 4.6$, $\tau_c = 50\ \mu s$, $k_B T/U_l = 0.05$ (blue), $k_B T/U_l = 0.13$ (red).

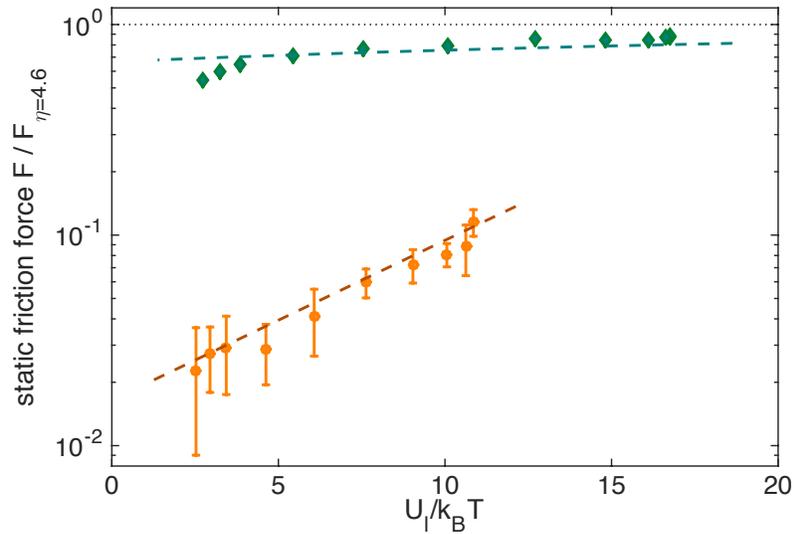

**Figure 3 | Thermolubricity for a single atom.** In the thermal drift regime $\tau_{th} \ll a/v$, the friction force is proportional to $\exp(U_l/k_B T)$, while it depends only weakly on temperature in the stick-slip regime[13] $\tau_{th} \gg a/v$. We vary temperature from $k_B T/U_l = 0.06$ to $k_B T/U_l = 0.4$, and show the friction force on a logarithmic scale against $1/T$. We fit data to the model $F/F_{\eta=4.6} = f \exp(cU_l/k_B T)$, where $f$ is a free parameter, and $c$ represents the fitted sensitivity to temperature and is close to unity in the thermal drift regime. For a high velocity ($v \approx 1$ mm/s) corresponding to the stick-slip regime (green), friction is almost constant for $U_l/k_B T \geq 5$, and the fit to the model in this temperature range (dashed green line) gives $c = 0.016$, i.e. very weak temperature dependence. For a low velocity ($v \approx 40$ μm/s) close to the regime of thermal drift (orange), the friction force is sensitive to temperature, and the fit to the model (red dashed line) gives $c = 0.17$. Experimental parameters are $\eta = 4.6$, $U_l/h = 20$ MHz, and $\tau_c \approx 50$ μs.

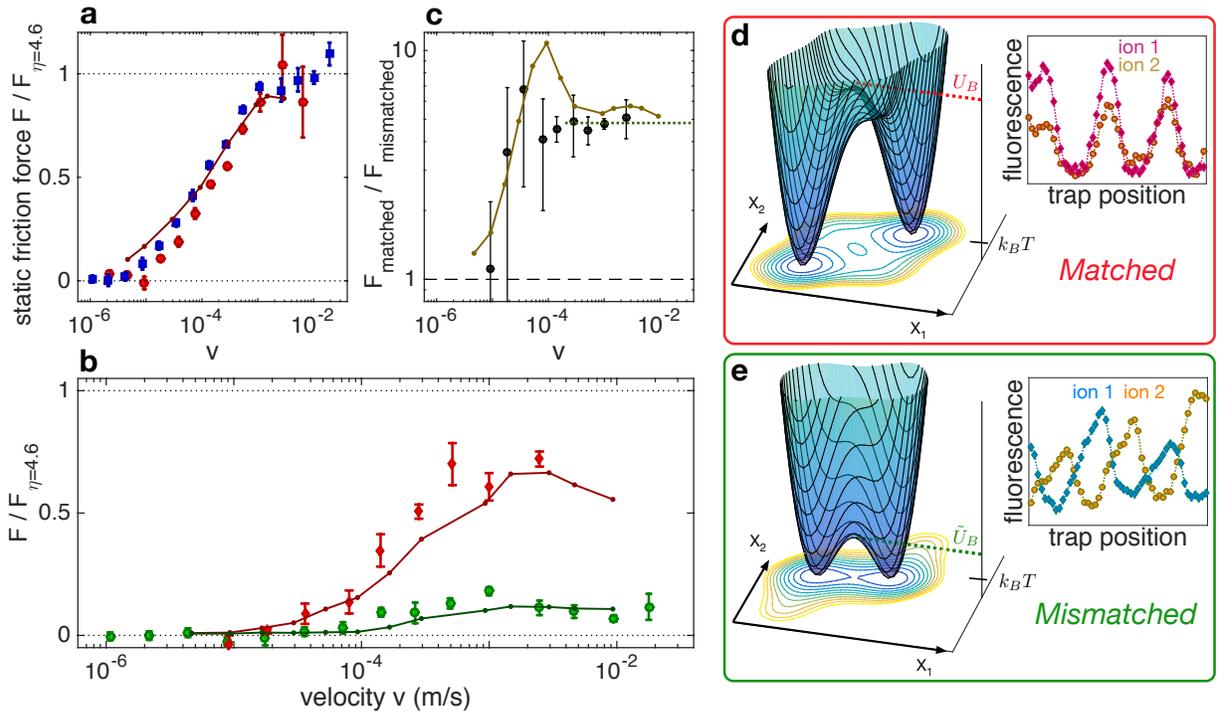

**Figure 4 | Structural and thermal lubricity of two atoms. a,b,c,** Velocity-dependence of the friction force for two ions for $\eta = 4.6$. **a,** In the matched case (red circles), where the ion spacing is an integer multiple of the lattice period $a$, for $k_BT/U_l = 0.055(10)$ the data agree with one ion at approximately the same temperature (blue squares), and reach a maximal value of $F_{\eta=4.6}$. **b,** In the mismatched case (green circles), where the two ions at their unperturbed position experience opposite forces by the optical lattice, the maximal friction is $\sim 0.15\, F_{\eta=4.6}$ at a temperature of $k_BT/U_l = 0.15(2)$. By comparison, in the matched case (red diamonds), at the same temperature of $k_BT/U_l = 0.15(3)$, the friction reaches a maximum of $\sim 0.7\, F_{\eta=4.6}$. Finite-temperature simulations (solid lines) are in good agreement with data for $\eta = 4.6$, $k_BT/U_l = 0.15$. **c,** The ratio of friction forces in the matched and mismatched cases (black circles, 3-point running average) is unity in the low-velocity thermal drift regime, and constant in the high-velocity stick-slip regime, where its value $\sim 4.8$ is mostly due to structural lubricity, in good agreement with Langevin simulations (solid gold line). **d,e,** Energy potential landscape for two interacting atoms. In the mismatched case (e), the energy barrier $\widetilde{U}_B$ between the wells is reduced by a factor of $\sim 3.7$ and the ions pass the barrier one at a time (inset), compared to the matched case $U_B$ (d) where the ions pass the barrier simultaneously (inset). At fixed $T$ for a single ion in the stick-slip regime, this barrier reduction $\widetilde{U}_B/U_B$ would lead to a thermal friction reduction of $\sim 1.4$ as can be inferred from the green data in Fig. 3. The expected total reduction of $3.7 \times 1.4 = 5.2$ is in good agreement with the observed reduction of $\sim 4.8$.

## Supplementary Information

### Measuring and controlling position and temperature $T$

Our laser-cooling scheme uses the optical lattice to couple the vibrational levels $n$ and $n-2$ of the ion's quantized motion in the optical lattice well[32]. The spatial dependence of this Raman coupling is such that the off-resonant transition $n \rightarrow n$, which on resonance would be stronger by two orders of the Lamb-Dicke factor $\eta_{LD}$ ($\eta_{LD} \sim 10\%$ for our system), increases from lattice node to lattice anti-node proportionally to the optical potential. The stronger this coupling is, the larger the scattered fluorescence, resulting in a position-dependent fluorescence signal, which, when time-resolved, amounts to sub-wavelength imaging of the ion's average trajectory.

To measure temperature, we make use of a near-detuned cavity standing-wave beam, which has a nearly identical spatial dependence around the ion position as the optical lattice, and which we pulse on for a short time so as not to change the temperature of the ion during measurement. If the ion is placed at the lattice node, its average fluorescence gives information about its average kinetic energy and temperature $T$. By taking the ratio of this signal to a signal taken by heating the ion to a temperature $k_B T_{HOT} \geq U_l$ (where the ion is completely delocalized and the fluorescence signal saturates), we obtain a measurement of temperature $k_B T / U_l$.

The temperature $T$ reached by the ion depends on the balance of cooling from laser light and heating from ambient electrical and optical noise. Deliberately increasing noise in the system leads to a higher temperature. In our experiments, we increase the ion's temperature with additional recoil heating from a near-detuned laser beam, allowing us to tune the temperature from $k_B T / U_l = 0.04$ in the absence of additional noise, to $k_B T / U_l \approx 1$, where our measurement saturates.

### Velocity-dependent temperature correction

In our system, due to our laser-cooling configuration, the temperature $T$ to which the ion thermalizes depends on its position relative to the optical potential. While conducting friction measurements for a slow enough drive velocity $v$, the ion thermalizes to a higher temperature before slipping to an adjacent lattice well. We can experimentally measure this velocity-dependent temperature of the ion before it slips, and we find it to be approximately $k_B T / U_l = 0.02$ of extra temperature per velocity decade below $v = 2\ mm/s$. This logarithmic correction is present in the numerical simulation curves presented in Figs. 2a,b and 4a,b, and in the theoretical curves from the analytical model presented in Fig. 2a,b.

### Measuring the recooling time $\tau_c$

The recooling time $\tau_c$ corresponds to the typical time our laser cooling scheme takes to dissipate the ion's energy down to its equilibrium value of $k_B T$. Because the ion's scattered fluorescence is directly proportional to its energy, $\tau_c$ can be obtained from a fit of the ion fluorescence versus time when the ion recools from a slip event. To measure this unambiguously, we initialize the position of the ion in one lattice well, by centering the electrical trap relative to it, and rapidly cause it to become unstable, by translating the electrical trap by one lattice spacing $a$, forcing the ion to instantaneously slip to the neighboring well (for $\eta \leq 4.6$). The ion acquires a large energy, and scatters a correspondingly large fluorescence, which decays exponentially towards $k_B T$ with a fitted typical time $\tau_c$.

**An analytical model for friction in the underdamped regime**

The thermal activation time $\tau_{th} = \tau_0 \exp(U/k_B T)$ is the typical time it takes the ion's energy to increase from the mean of its energy distribution $k_B T$ to the tail of its energy distribution $U \gg k_B T$. In a complementary way, the recooling time $\tau_c$ is the typical time it takes the ion's energy to decrease from the tail of its energy distribution $U \gg k_B T$ to the mean of its energy distribution $k_B T$. In the regime $a/v \sim \tau_{th}$ where the transport time $a/v$ competes with the thermal activation time $\tau_{th}$ for causing the ion to pass the energy barrier $U$, the friction force grows with velocity[3,21] as
$F/F_{max} = 1 - \left(\frac{3}{2\sqrt{2}} \frac{k_B T}{U_l} \log\left(\frac{v_{th}}{v}\right)\right)^{2/3}$, where $v_{th} = \frac{1}{2\sqrt{2\pi}} \frac{k_B T}{U_l} \frac{\eta}{\eta+1} \frac{a}{\tau_0}$. It is then natural to extend this relationship to the reverse process: in the regime $a/v \sim \tau_c$ where the transport time $a/v$ competes with the recooling time $\tau_c$ for an ion which passed the energy barrier $U$, the friction force will decrease with velocity as $F/F_{max} = 1 - \left(\frac{3}{2\sqrt{2}} \frac{k_B T}{U_l} \log\left(\frac{v}{v_c}\right)\right)^{2/3}$, where we take $v_c = \frac{a}{\tau_c}$ as the simplest model which agrees well with the data.

**Numerical simulations of stick-slip friction**

We follow standard numerical methods[11,21] for calculating the mean friction force experienced by a single particle in a periodic potential under the influence of an external shear force and a fluctuating force. We calculate the average trajectory $\langle x(t) \rangle$ of the particle over multiple integrations of the Langevin equation of motion:
$$m\ddot{x} + m\gamma\dot{x} + K(x - vt) + \frac{\pi U_l}{a}\sin\left(\frac{\pi x}{a}\right) = \xi(t),$$
where the fluctuating force $\xi(t)$ satisfies the fluctuation-dissipation theorem which relates its magnitude to the temperature $T$ and damping coefficient $m\gamma$: $\langle \xi(t)\xi(t')\rangle = 2m\gamma k_B T \delta(t - t')$, where $\gamma = \tau_c^{-1}$. The friction force is the value of the external force on the particle when the particle slips, which reduces to $F \approx \frac{1}{\pi}\frac{(\eta-1)^2}{\eta}Ka$, where $\eta = \frac{2\pi^2 U_l}{Ka^2} = \frac{\omega_l^2}{\omega_0^2}$. This value is then averaged across integrations of the Langevin equation to obtain the mean friction force presented in the simulation curves throughout this paper.